\documentclass[journal=nalefd,manuscript=letter]{achemso}

\usepackage[version=3]{mhchem} 
\usepackage{graphicx,epstopdf}
\usepackage{upgreek}

\author{Yonathan Anahory}
\altaffiliation{These authors contributed equally to this work.}
\email{yonathan.anahory@weizmann.ac.il}

\author{Jonathan Reiner}
\altaffiliation{These authors contributed equally to this work.}
\email{jonathan.reiner@weizmann.ac.il}

\author{Lior Embon}
\author{Dorri Halbertal}
\author{Anton Yakovenko}
\author{Yuri Myasoedov}

\author{Michael L. Rappaport}
\affiliation[Weizmann Institute of Science]
{Weizmann Institute of Science, Department of Condensed Matter Physics, 76100 Rehovot, Israel}

\author{Martin E. Huber}
\affiliation[University of Colorado Denver]
{Department of Physics, University of Colorado Denver, Denver, CO 80027, USA.}

\author{Eli Zeldov}
\email{eli.zeldov@weizmann.ac.il}
\affiliation[Weizmann Institute of Science]
{Weizmann Institute of Science, Department of Condensed Matter Physics, 76100 Rehovot, Israel}

\title
  {A three-junction SQUID-on-tip with tunable in-plane and out-of-plane magnetic field sensitivity}
	
\begin{document}

\begin{abstract}
Nanoscale superconducting quantum interference devices (SQUIDs) demonstrate record sensitivities to small magnetic moments, but are typically sensitive only to the field component that is normal to the plane of the SQUID and out-of-plane with respect to the scanned surface. We report on a nanoscale three-junction Pb SQUID which is fabricated on the apex of a sharp tip. Because of its three-dimensional structure, it exhibits a unique tunable sensitivity to both in-plane and out-of-plane fields.  We analyze the two-dimensional interference pattern from both numerical and experimental points of view. This device is integrated into a scanning microscope and its ability to independently measure the different components of the magnetic field with outstanding spin sensitivity better than $5\ \frac{\mu_B}{\mathrm{Hz}^{1/2}}$ is demonstrated. This highlights its potential as a local probe of nanoscale magnetic structures.\vspace*{0.25in}

\end{abstract}

The rich and diverse research in the field of nanoscale physics has given rise to exploration of many interesting phenomena in which magnetic interactions play an important role. This environment creates a need for precise and versatile magnetic characterization, prompting the development of magnetic imaging techniques that concentrate on imaging small magnetic moments with high spatial resolution \cite{Bending1999,Grinolds2013,Degen2011,Rugar2004,Tang2011}. Superconducting quantum interference devices (SQUIDs) have traditionally been an important tool due to their high magnetic sensitivity, particularly following recent advancement in nano-SQUID fabrication \cite{Finkler2010,Finkler2012,Hasselbach2002,Ronzani2013,Wolbing2013,Lam2011,Foley2009}, SQUID measurement techniques \cite{Kirtley1995,Nagel2011,Troeman2007,Kirtley2010, Hao2008,Levenson-Falk2013,Clarke2006} and scanning SQUID microscopy \cite{Koshnick2008,Hykel2014,Vasyukov2013}.

Scanning SQUID microscopy is predominately sensitive to the magnetic field component that is out-of-plane with respect to the scanned surface. However, in several applications, such as the study of local current distributions\cite{Sochnikov2013}, current-carrying edge states\cite{Bid2010}, transport in surface states\cite{Nowack2013}, spin-polarized currents\cite{Tiemann2012}, and detection of magnetic moments with in-plane polarization \cite{Kuemmeth2008,Kalisky2012}, it is the in-plane component of the magnetic field that provides the most local information about the magnetic moments. This limitation was previously addressed by fabrication of a three-dimensional pickup loop that can be oriented to measure the in-plane component\cite{Romans2010}. This geometry complicates significantly the experimental setup and needs to be reoriented to measure any other component.

Lately, a new method for fabrication of nano-SQUIDs on the apex of a sharp quartz tip was developed that eliminates the need for complex lithographic processes, and allows scanning with the SQUID-on-tip (SOT) within several nanometers of the scanned surface \cite{Finkler2010,Finkler2012,Vasyukov2013}. This device demonstrated a record magnetic moment sensitivity of $0.38\ \mu_B/\mathrm{Hz}^{1/2}$ \cite{Vasyukov2013}. 

In this work, we report a novel device: a three-junction SQUID-on-tip (3JSOT) in a three-dimensional configuration. Although SQUIDs with three junctions in parallel have been used for a number of purposes \cite{Chiarello2008,Martinez-Perez2013,Ronzani2014}, a device with three Josephson junctions in parallel has not been previously reported in the context of magnetic imaging. The 3JSOT utilizes the benefits of the SOT, yet it is sensitive to both the in-plane and the out-of-plane components of the magnetic field. This sensitivity can be tuned \textit{in-situ} to measure either of these orthogonal components.

\begin{figure}
			\begin{center}
				\includegraphics[width=0.8\textwidth]{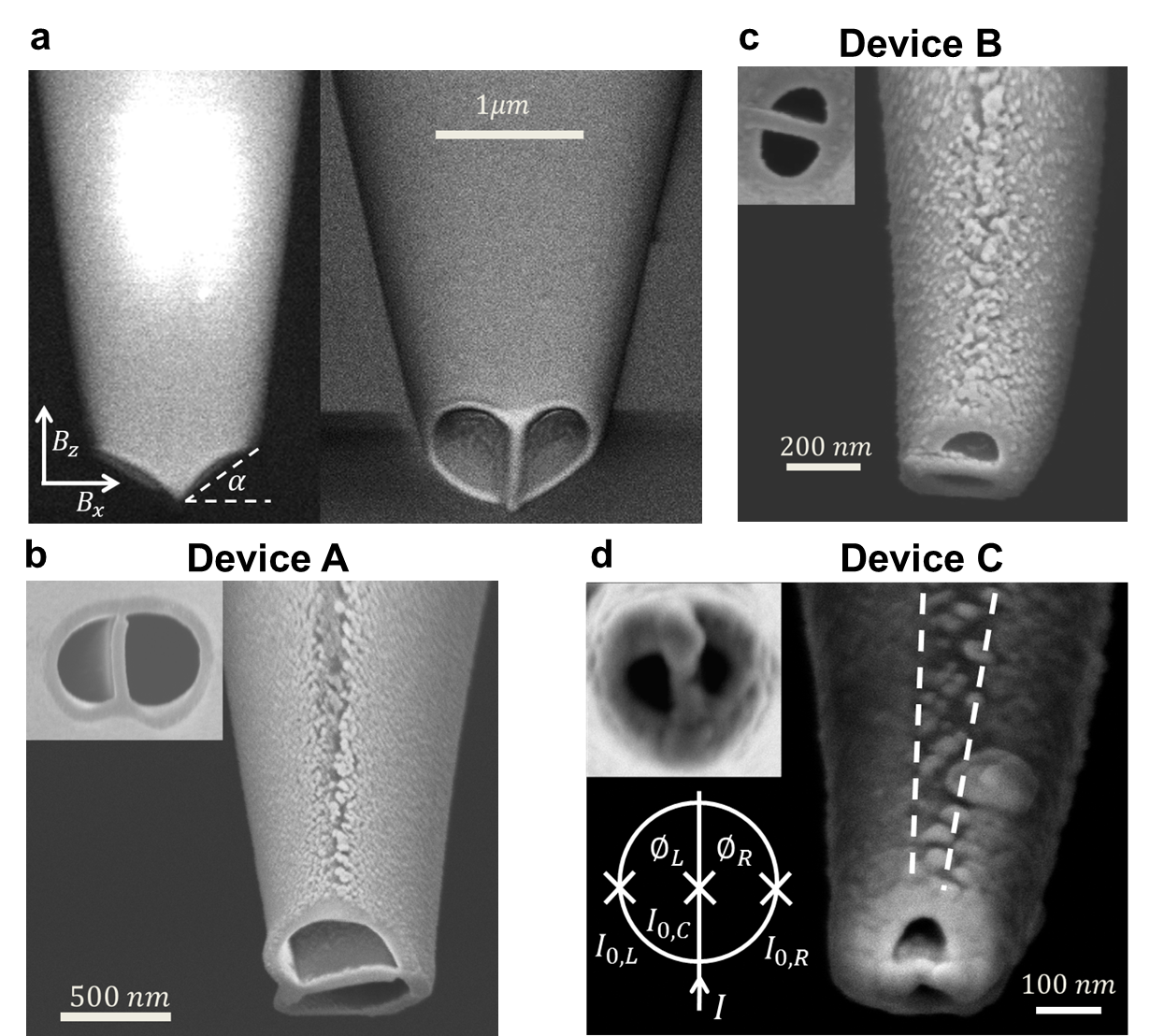}
				\caption[]{\label{FIBandSEM} 
				{\bf Scanning electron microscope images of a tapered $\theta$-tip and 3JSOT devices.} (a) Images of a bare borosilicate $\theta$-tip after FIB patterning. The magnetic field orientation and the tapering angle $\alpha$ are denoted.  (b) A 3JSOT (device A) with effective loop area of $A\cong 0.45\ \upmu \mathrm{m}^2$ ($A_L\cong 0.27\ \upmu \mathrm{m}^2$ and $A_R\cong 0.18\ \upmu \mathrm{m}^2$). (c) A 3JSOT (device B) with effective loop area of $A\cong 0.077\ \upmu \mathrm{m}^2$ ($A_L\cong 0.04$ and $A_R\cong 0.037\ \upmu \mathrm{m}^2$). (d) A 3JSOT (device C) with effective loop area of $A\cong 0.042\ \upmu \mathrm{m}^2$ ($A_L\cong A_R\cong 0.021\ \upmu \mathrm{m}^2$), later used for noise characterization (Fig.\ \ref{Spectra}) and imaging (Figs. \ref{2dscans} and \ref{profiles}). The gap that separates the two evaporated Pb leads is marked by the dashed lines, and the inset shows a schematic of a 3JSOT.}
			\end{center}
		\end{figure}

Since the SQUID response is fundamentally a function of magnetic flux, which is a scalar quantity, an additional degree of freedom must be introduced in order to induce independent sensitivity to the different components of the magnetic field. 
Whereas the conventional SOT is fabricated by pulling a pipette with a circular cross section to sub-micron dimensions, a 3JSOT is based on a pipette  with $\theta$-shaped cross section (Fig. \ref{FIBandSEM}). A borosilicate capillary (OD=1 mm, ID=0.7 mm) with a central partition is heated with a laser and subsequently pulled to form a sharp tip. The final size of the apex is controlled by the exact pulling parameters and can have an overall diameter as small as 150 nm. When a superconductor is deposited by the self-aligned deposition technique described in detail in Refs. \citenum{Finkler2010} and \citenum{Vasyukov2013}, this new geometry results in the formation of a double-loop, triple-junction SQUID on the apex of the tip. Two junctions reside along the circumference; the third junction, common to both loops, is formed on the central partition (insets in Fig. \ref{FIBandSEM} b-d). Consequently, the properties of this device are determined by two parameters, $\Phi_L$ and $\Phi_R$ - the magnetic flux threading each loop, left and right, respectively. The fact that there are two fluxes, rather than just one, ultimately provides the required additional degree of freedom.  

As long as the 3JSOT geometry remains essentially two-dimensional, i.e., the loops are in the same plane, only one component of the field produces flux that threads the loops. This restriction can be removed by introducing a three-dimensional configuration in the following manner. Prior to deposition of the superconductor ($20$ nm-thick Pb layer), the tip is milled by a focused ion beam (FIB) to a ``V'' shape at a prescribed angle $\alpha$. After milling, the 3JSOT loops are at an oblique angle with respect to each other, and this allows flux to couple from both the out-of-plane component $B_{z}$ and the in-plane component $B_{x}$ (Fig.\ \ref{FIBandSEM}a).

Recently, the SOT has matured into a reliable and effective tool, manifested in high fabrication yield ($\sim 90\%$). The additional complication in 3JSOT fabrication comes from the FIB patterning. Using a scanning electron microscope (SEM) allows post-selection of the most suitable tips, and therefore the high production yield remains almost unaffected. To meet the sensitivity requirements of a specific experiment, several 3JSOTs may need to be characterized. Pb based SOTs, in general, can endure approximately a day when exposed to atmosphere, but can last for weeks and can even be reused when kept under vacuum.       

\begin{figure}
				\begin{center}
					\includegraphics[width=0.8\textwidth]{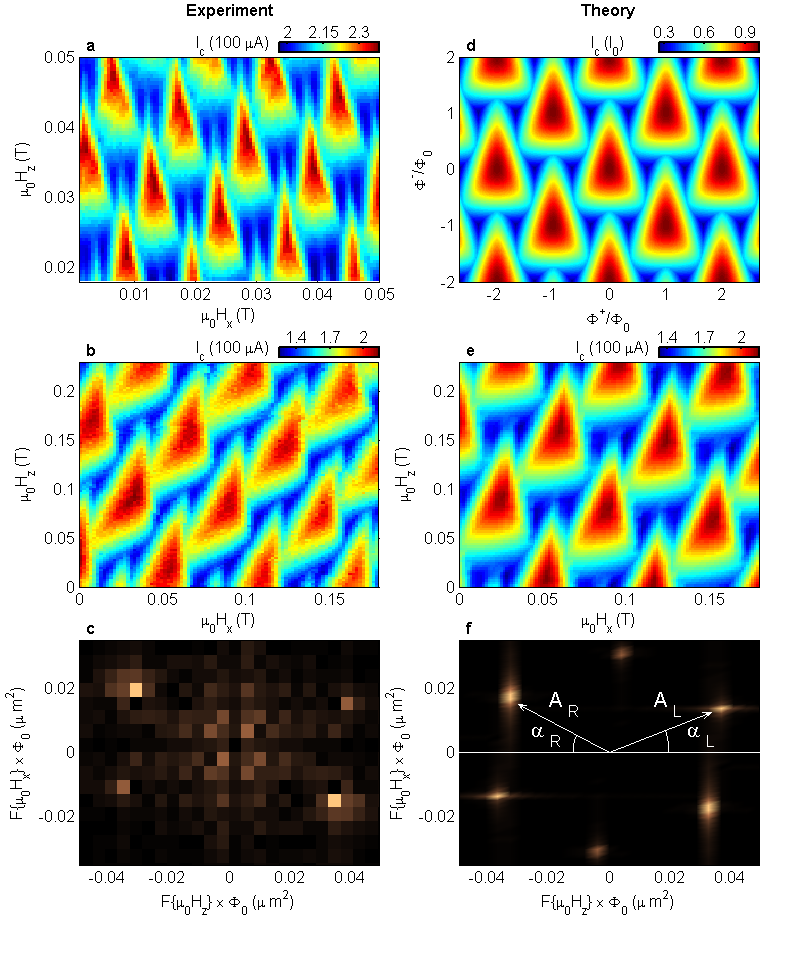}
					\caption[]{\label{Results} {\bf 3JSOT interference patterns.} (a) Measured $I_c(H_{x},H_{z})$ of device A (Fig.\ \ref{FIBandSEM}b). The orientation of the periodic lattice is determined only by the geometrical structure of the 3JSOT, whereas the shape of the individual triangular peaks depends also on the critical currents and the inductances of the junctions. (b) Measured $I_c(H_{x},H_{z})$ of device B.  (c) The FFT of (b), from which the geometrical parameters are extracted: $A_L=0.04\ \upmu \mathrm{m}^2$, $A_R=0.037\ \upmu \mathrm{m}^2$, $\alpha_L=19^{\circ}$, and $\alpha_R=27^{\circ}$. (d) Simulation of $I_c\left(\Phi^{+},\Phi^{-}\right)$ for $I_{0,L}=I_{0.C}=I_{0,R}=I_{0}/3=1.65\ \mathrm{\upmu A}$ and $L_L=L_C=L_R=0.55\ \mathrm{nH}$. (e) A numerical fit to (b). (f) FFT of (e). The arrows indicate the locations of the maxima and the quantities that can be derived from them.}
				\end{center}
	\end{figure}

Figures \ref{Results} a and b show the measured interference pattern of the critical current, $I_c(H_{x},H_{z})$ for two different devices, A and B (corresponding to the SEM micrographs in Fig.\ \ref{FIBandSEM} b and c).  In contrast to conventional two-junction SQUIDs that display 1D interference patterns $I_c(H)$, the 3JSOTs exhibit 2D interference patterns, $I_c(H_{x},H_{z})$. This unique 2D nature of the interference patterns is the feature enabling the use of the 3JSOT as a 2-axis magnetometer, since $H_{x}$ and $H_{z}$ can be tuned so that the response to small field variations is effectively decoupled for the different components, as will be discussed below. Here we use $H$ to indicate the externally applied magnetic field used to bias the device, where $B$ indicates the measured field, which is affected by the sample's magnetization.

The 2D interference pattern can be understood by applying the DC Josephson relation $I=I_0 \sin{(\delta)}$ to the three-junction device case, where $\delta$ is the superconducting order parameter phase difference across a junction. The total current $I_{total}$ flowing through the 3JSOT is determined by the flux threading the two loops, and can be written in terms of:
			\begin{align} 
			I_{total}=I_{0,L}\sin{\delta_L}+I_{0,C}\sin{\delta_C}+I_{0,R}\sin{\delta_R},	\label{TotalCurrent}\\ 			
			\delta_{L}-\delta_{C}=\frac{2\pi}{\Phi_0}\left(\Phi_{L}-L_LI_{0,L}\sin{\delta_L}+L_CI_{0,C}\sin{\delta_C}\right), \label{PhaseFlux1}\\				
			\delta_{C}-\delta_{R}=\frac{2\pi}{\Phi_0}\left(\Phi_{R}-L_CI_{0,C}\sin{\delta_C}+L_RI_{0,R}\sin{\delta_R}\right),  \label{PhaseFlux2}
			\end{align}
			 
Here $\Phi_0=2.07\times10^{-15}\ \mathrm{Wb}$ is the flux quantum, $\Phi_{L}$ and $\Phi_{R}$ are the external applied fluxes in the corresponding loop, $I_0$ is the critical current, $L$ is the self-inductance (see the schematic in Fig.\ \ref{FIBandSEM}d inset) and the subscripts L, R and C denote the left, right and central junctions, respectively. In this analysis we neglect the mutual inductance since, due to the small size of the device, the inductances are dominated by the kinetic inductance \cite{Finkler2010,Vasyukov2013}.
			Upon setting $\Phi_{L}$ and $\Phi_{R}$ as independent parameters, we can solve Eqs. (1)-(3) numerically and derive $I_c\left(\Phi_{L},\Phi_{R}\right)$. For our purpose, it is useful to express $\Phi_L$ and $\Phi_R$ in terms of the total flux in the 3JSOT $\Phi^+=\Phi_L+\Phi_R$ and the flux difference $\Phi^-=\Phi_L-\Phi_R$. The resulting calculated $I_c\left(\Phi^{+},\Phi^{-}\right)$ in Fig. \ref{Results}d, shows a periodic square lattice (rotated by $45^{\circ}$) of triangular peaks. A similar interference pattern has been observed in Ref. \citenum{Chiarello2008}, where different fluxes were applied to different loops of a three-junction Josephson device. For this specific case, of a device with identical critical currents and inductances for all three junctions, the triangular peaks are symmetric. The shape of the peaks can be deformed, skewed or uniformly shifted for arbitrary junction parameters; however, the lattice structure of $I_c(\Phi^+,\Phi^-)$ remains square for any set of parameters.
  
In order to quantitatively compare the numerical to the experimental results, one must take into account the actual three-dimensional geometry to express $I_c$ as a function of $H_{x}$ and $H_{z}$, rather than of $\Phi^+$ and $\Phi^-$. For a 3JSOT with tapering angles $\alpha_{L}$ and $\alpha_{R}$ and effective loop areas $A_{L}$ and $A_{R}$, the fluxes are associated with the fields by the following transformation, assuming uniform fields: 			
			\begin{align} \label{eq:FluxToField}
			\begin{pmatrix}
			\Phi^+ \\
			\Phi^-
			\end{pmatrix}
			=
			\begin{pmatrix}
			A_L\cos{\alpha_L}+A_R\cos{\alpha_R} & A_L\sin{\alpha_L}-A_R\sin{\alpha_R} \\
			A_L\cos{\alpha_L}-A_R\cos{\alpha_R} & A_L\sin{\alpha_L}+A_R\sin{\alpha_R}
			\end{pmatrix}
			\begin{pmatrix}
			\mu_{0} H_{z} \\
			\mu_{0} H_{x}
			\end{pmatrix},
			\end{align}
			where $\mu_{0}$ is the vacuum permeability. In the case of a symmetric 3JSOT ($\alpha_R=\alpha_L=\alpha$, $A_R=A_L=A/2$), the off-diagonal terms vanish and the two fluxes are simply given by $\Phi^+ = \mu_{0} H_z (A \cos\alpha)$ and $\Phi^- = \mu_{0} H_x (A\sin\alpha)$. Namely, $H_z$ determines the total flux $\Phi^+$ in the 3JSOT, while $H_x$ governs the flux difference $\Phi^-$ between the loops (see Fig. \ref{FIBandSEM}a). Considering the inverse transformation of (\ref{eq:FluxToField}) we can compute how the interference pattern $I_c(\Phi^+,\Phi^-)$ is modified when converting its arguments from fluxes to fields, as a function of the geometrical parameters. These parameters have a pronounced impact on the pattern, as they not only affect the shape of the individual peaks but also distort the original square lattice, changing its periodicity and directionality (compare Fig.\ \ref{Results} d and e). This fact implies that the geometrical parameters of a 3JSOT can be extracted independently from the microscopic parameters by performing 2D fast Fourier transform (FFT) of the measured $I_c(H_{x},H_{z})$, since the first-order peaks of the FFT are robust against the deformation of the individual triangles. This process is equivalent to the determination of the effective area of a two-junction SQUID from its oscillation period, as verified self-consistently by performing an FFT on our simulations (Fig.\ \ref{Results}f). As an example, the FFT of the interference pattern of device B (Fig. \ref{FIBandSEM}c) is shown in Fig.\ \ref{Results}c, with derived values of $A_L=0.04\ \mu \mathrm{m}^2$, $A_R=0.037\ \mu \mathrm{m}^2$, $\alpha_L=19^{\circ}$, and $\alpha_R=27^{\circ}$, consistent with the SEM images.
			
Having the geometric parameters, we can evaluate the microscopic parameters of the junctions by numerical fitting (Fig. \ref{Results}e), resulting in critical currents $I_{0,L} \cong 36\ \upmu \mathrm{A}$, $I_{0,C} \cong 95\ \upmu \mathrm{A}$, and $I_{0,R} \cong 77\ \upmu \mathrm{A}$, and inductances $L_{L} \cong 20\ \mathrm{pH}$, $L_{C} \cong 10\ \mathrm{pH}$, and$L_{R} \cong 15\ \mathrm{pH}$. Our findings indicate that the effective width and thickness of the central Dayem bridge is larger, naturally giving rise to higher critical current. It is interesting to note the inverse correlation between the critical current and the inductance of each junction. In our small devices, the inductances are governed by the kinetic inductance, $L_k=\frac{\mu_0\lambda_{L}^{2}l}{wt}$, where $\lambda_{L}$ is the London penetration depth, and $l$, $t$, and $w$ are the length, thickness and width of the bridge, respectively. Thus, a wider and thicker bridge will have a larger $I_c$ and a lower kinetic inductance. 

The 2D periodic structure of $I_c(H_{x},H_{z})$ of the 3JSOT allows realization of a 2-axis vector magnetometer with tunable sensitivity to in-plane and out-of-plane fields. One of the important applications of such a device is scanning probe microscopy of weak local magnetic fields arising from magnetic nanoparticles in the presence of an external magnetic field. For this application, we voltage-bias the device and measure the current $I$ flowing through it, as described in Refs. \citenum{Finkler2010} and \citenum{Vasyukov2013}. We define the in-plane and out-of-plane response functions as $r_{x}= \partial I / \partial B_{x}$ and $r_{z}= \partial I / \partial B_{z}$, respectively. For a convenient use of the 3JSOT as a 2-axis vector magnetometer, good decoupling of the response functions is desirable. This outcome can be readily achieved by field-biasing the 3JSOT to regions of $I_c(H_{x},H_{z})$ where one of the response functions is large and the other vanishes, i.e., where the contour lines of the $I_c(H_{x},H_{z})$ plot are parallel to one of the axes. More generally, one can perform the measurements at two nearby field-biased working points (WPs) with known (and different) response to applied $H_{x}$ and $H_{z}$ and then reconstruct the in-plane and out-of-plane field components from a linear combination of the two mixed signals.  

\begin{figure}
				\begin{center}
					\includegraphics[width=0.5\textwidth]{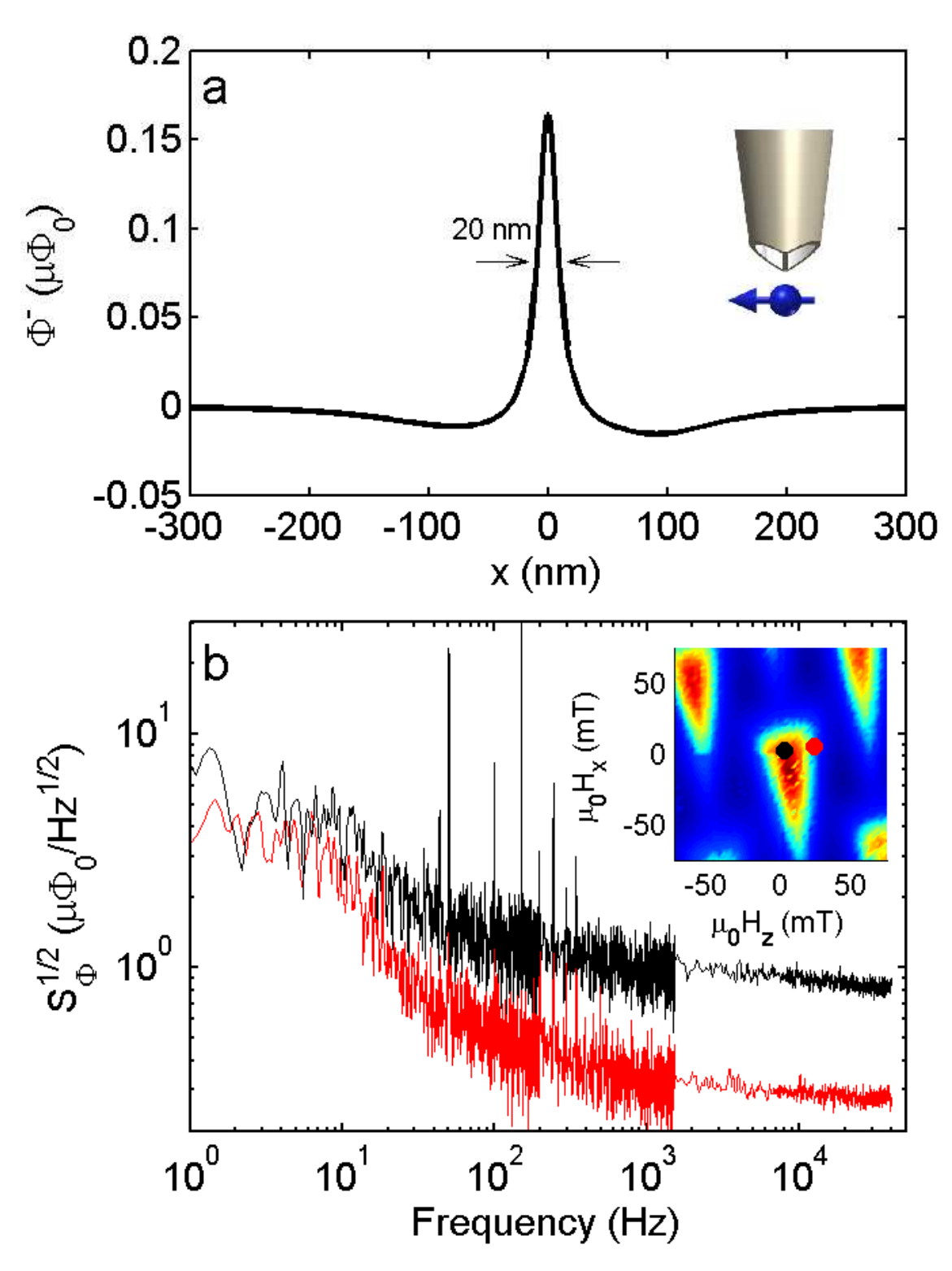}
					\caption[]{\label{Spectra} {\bf Flux response and sensitivity of a 3JSOT.} (a) Calculated flux coupling $\Phi^-$ upon scanning device C over an in-plane magnetic dipole of $1\ \mu_B$ at a height of 10 nm, as illustrated in the inset.  (b) Measured flux noise spectral densities $S_{\Phi}^{z\ 1/2}$ and $S_{\Phi}^{x\ 1/2}$ at the two WPs sensitive to $B_{z}$ (red, $\mu_{0} H_{z}=24.2\ \mathrm{mT}$, $\mu_{0} H_{x}=5.0\ \mathrm{mT}$ ), and $B_{x}$ (black, $\mu_{0} H_{z}=2.5\ \mathrm{mT}$, $\mu_{0} H_{x}=2.3\ \mathrm{mT}$). Inset: measured $I_c(H_{x},H_{z})$ around the working regions used for scanning. The color scale is $180\ \upmu \mathrm{A}$ (blue) to $240\ \upmu \mathrm{A}$ (red). The red (black) dot shows the WP selected to measure the spectra with high decoupling ratio for $B_{z}$ ($B_{x}$).}
				\end{center}
			\end{figure}

To determine the sensitivity of our devices, we conducted a systematic noise characterization of a 3JSOT (device C), which was later used for scanning. Figure \ \ref{Spectra}b shows the spectral flux noise density at the two indicated WPs (inset) that have a high decoupling ratio: at the $B_{x}$-sensitive point (black) the decoupling ratio is $r_{x}/r_{z}\sim 10$ while at the $B_{z}$-sensitive point (red) $r_{z}/r_{x }\sim 30$. Note that Fig \ref{Spectra}b indicates that numerous WPs with high decoupling ratio can be chosen for various needs as shown below. Both spectra display $1/f$ noise at frequencies of up to about 1 kHz, followed by white noise of $S_{\Phi}^{x\ 1/2}=800\ \mathrm{n}\Phi_0/\mathrm{Hz}^{1/2}$ (field noise $S_B^{x\ 1/2}=70\ \mathrm{nT}/\mathrm{Hz}^{1/2}$) and $S_{\Phi}^{z\ 1/2}=280\ \mathrm{n}\Phi_0/\mathrm{Hz}^{1/2}$ ($S_B^{z\ 1/2}=20\ \mathrm{nT}/\mathrm{Hz}^{1/2}$) at the $B_{x}$ and $B_{z}$-sensitive points, respectively. The field sensitivity is calculated by dividing the signal noise density spectra by the response functions of the 3JSOT, $r_{x}$ and $r_{z}$. The flux sensitivity was calculated from the field sensitivity using Eq. \ref{eq:FluxToField}. 

The in-plane field sensitivity of the 3JSOT is highly advantageous for the study of magnetic nanoparticles with in-plane magnetization. Figure \ref{Spectra}a shows the calculated flux $\Phi^{-}$ coupled to the 3JSOT upon scanning $10$ nm above a single electron with in-plane spin orientation using the geometric parameters of device C. The advantage of the 3JSOT is that the spatial dependence of the flux coupling has a very sharp local peak directly above the spin with high resolution of 20 nm. This situation is in contrast to conventional SQUIDs that are only sensitive to $B_z$ that would display an extended non-local feature when imaging in-plane oriented spins\cite{Vasyukov2013}. Using this calculated flux coupling, our measured flux noise translates into spin noise of $4.9\ \mu_B/\mathrm{Hz}^{1/2}$ for in-plane spins. 
			
\begin{figure}
				\begin{center}
					\includegraphics[width=1\textwidth]{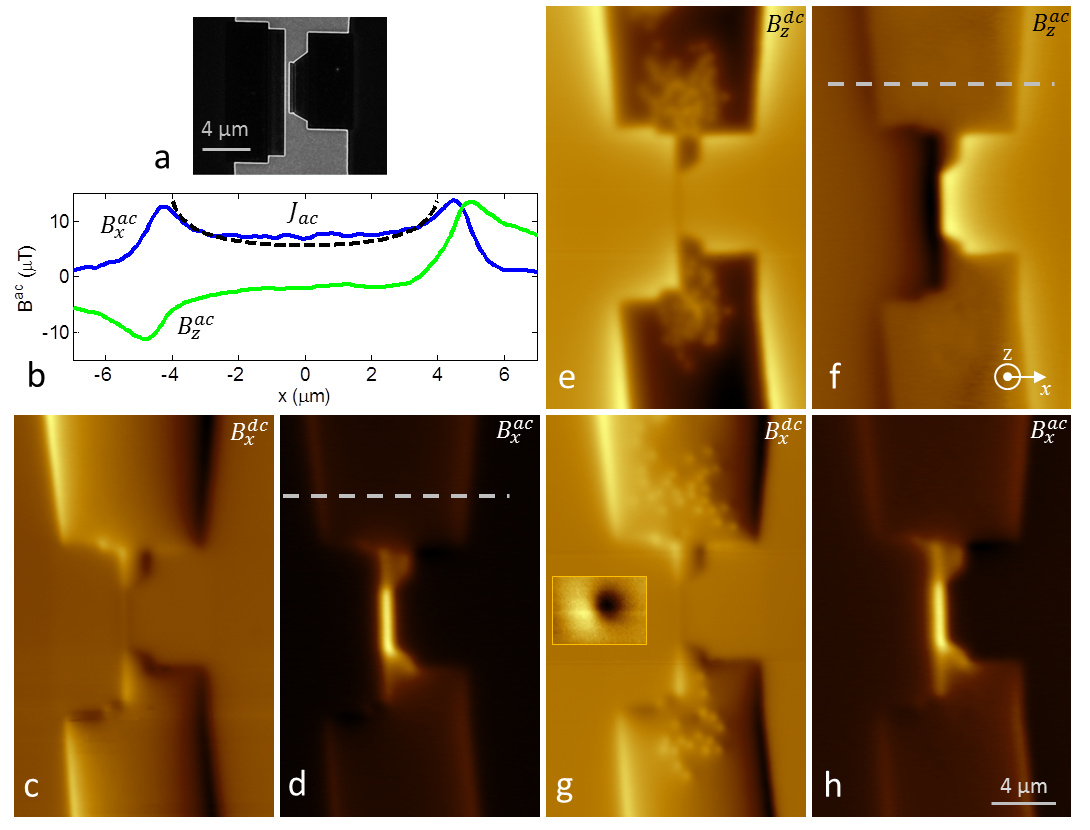}
					\caption[h!]{\label{2dscans} {\bf Images of the magnetic field in a patterned Pb film} (a) $12 \times 15\ \upmu \mathrm{m^2}$ SEM image of a Pb film with a $0.35\times4\ \upmu \mathrm{m^2}$ nanowire. (b) Measured $B^{ac}_x$ (blue) and $B^{ac}_z$ (green) along the dashed lines in (d) and (f) and theoretical current density distribution (black dashed).  (c)-(h) 3JSOT scanning microscopy images ($16 \times 25\ \upmu \mathrm{m^2}$) with an AC current of $150\ \upmu \mathrm{A}$ at 10.4 kHz applied to the sample; (c) $B^{dc}_x$ and (d) $B^{ac}_x$ were acquired at $WP_x$ ($\mu_{0} H_{z} = 0\ \mathrm{mT}$ and $\mu_{0} H_{x} = 32\ \mathrm{mT}$), when the sample is in the complete Meissner state; (e) $B^{dc}_z$ and (f) $B^{ac}_z$ were acquired at $WP_z$ ($\mu_{0} H_{z}=-3.5\ \mathrm{mT}$ and $\mu_{0} H_{x}=-10\ \mathrm{mT}$) when the sample is in the mixed state;  (g) $B^{dc}_x$ and (h) $B^{ac}_x$ acquired at $WP_x$. Under these conditions, the sample should be free of vortices. However, this frame was taken after (e)-(f), retaining the previously induced vortices in the sample. The inset in (g) is a $3\times4.2\ \upmu \mathrm{m^2}$ $B_x^{dc}$ image of a single vortex. For the $B^{dc}$ images, the dark-to-bright full scales are as follows : (c) $10.8\ \mathrm{mT}$,  (e) $14.2\ \mathrm{mT}$  (g) $14.0\ \mathrm{mT}$. For the $B^{ac}$ images, the dark-to-bright range are : (d) $-1.58$ to $69.5\ \upmu \mathrm{T}$,(f) $-30.5$ to $35.9\ \upmu \mathrm{T}$,(h) $- 6.80$ to $56.3\ \upmu \mathrm{T}$.}  
					\end{center}
			\end{figure}

Device C was integrated into an in-house-built scanning probe microscope, operated at 4.2 K \cite{Vasyukov2013}. As a proof of concept, we measured both components of the field generated by currents in a patterned Pb film. A 100 nm-thick Pb film coated \textit{in-situ} with a 10 nm Ge layer was deposited on a Si substrate at $\sim 77\ \mathrm{K}$. A combination of optical lithography, followed by FIB patterning was used to obtain a 350 nm-wide and $4\ \upmu \mathrm{m}$-long nanowire, as shown in Fig. \ref{2dscans}a. 

An AC current of $150\ \upmu \mathrm{A}$ at 10.4 kHz was applied to the sample and the corresponding AC and DC magnetic fields, $B^{ac}$ and $B^{dc}$, respectively, were imaged simultaneously by the scanning 3JSOT, as shown in Fig. \ref{2dscans}. By tuning the applied DC magnetic field $H$ to appropriate sensitive WPs, we can measure the local $B_z$ and $B_x$ field variations ($WP_x$: $\mu_{0} H_z = 0\ \mathrm{mT}$ and $\mu_{0} H_x = 32\ \mathrm{mT}$;  $WP_z$: $\mu_{0} H_z = -3.5\ \mathrm{mT}$ and $\mu_{0} H_x = -10\ \mathrm{mT}$). $WP_x$ was chosen to allow the sample to be in the Meissner state upon initial cooling, as shown in Fig. \ref{2dscans}c. On changing the applied field to $WP_z$, some vortices penetrate the wide regions of the Pb film, as seen in Fig. \ref{2dscans}e, while the narrow regions and the central nanowire remain vortex-free. Since the applied AC transport current is much smaller than the critical current ($I_c > 10$ mA), the vortices do not move and hence the distribution of the transport current across the sample has the same Meissner-like form with and without vortices \cite{Zeldov1994}. We can thus directly compare the $B_x^{ac}$ and $B_z^{ac}$ field components shown in Fig. \ref{2dscans} d and f that originate from the same AC current distribution. We first analyze the fields in the wide regions of the sample. Figure \ref{2dscans}b shows the field profiles along the dashed line in Fig. \ref{2dscans} d and f. In a thin strip of width $2w$, the distribution of the transport current in the Meissner state is given by $J_y(x)=I/(\pi\sqrt{w^2-x^2})$, where $J$ is the sheet current density and $I$ is the total applied current \cite{Zeldov1994}. This current distribution $J_y(x)$ is shown by the dashed line in Fig. \ref{2dscans}b. The comparison between $J_y(x)$ and $B_x^{ac}$ and $B_z^{ac}$ highlights the advantage of measuring $B_x$. The contribution to $B_x(x_0)$ due to a current element $J_y$ of width $\mathrm{d}x$ at $x_0$ is simply given by $\mu_0J_y(x_0)\mathrm{d}x/2$. Thus, at close proximity to the surface, $B_x(x)$ provides a direct measure of $J_y(x)$. In contrast, the contribution of this current element to $B_z(x_0)$ vanishes, resulting in a highly non-local dependence of $B_z(x)$ on $J_y(x)$. Thus, the current distribution can be extracted from the in-plane field in a straightforward manner, whereas existing scanning probe techniques, which usually measure only the out-of-plane field, require the use of elaborate non-local inversion techniques \cite{Schuster1995,Dinner2005}.

The locality of the information provided by $B_x$ is also visible when inspecting $B_x^{ac}$ and $B_z^{ac}$ across the nanowire in Fig. \ref{2dscans} d and f. While Fig. \ref{2dscans}d shows a very sharp signal along the nanowire reflecting the high current density there, Fig. \ref{2dscans}f shows a broad signal outside the nanowire, where no current flows. A similar conclusion is drawn by comparing the DC signals $B_x^{dc}$ and $B_z^{dc}$ in Fig. \ref{2dscans} c and e. In the Meissner state, the DC shielding currents flow in opposite directions along the two edges of a strip. The $B_x^{dc}$ image in Fig. \ref{2dscans}c directly shows these currents as a bright and dark signal on the opposite edges inside the wide strips of the sample, while outside the strips the signal drops sharply to zero. In Fig. \ref{2dscans}e, however, the Meissner regions show zero signal, even though shielding currents do flow there, while the regions outside the edges show an enhanced bright signal, where no current flows. For the study of vortices, in contrast, $B^{dc}_z$ provides a sharp local peak at the vortex center, while in the $B^{dc}_x$ image, the vortices appear as dipoles with zero signal at their centers, as shown in the inset of Fig. \ref{2dscans}g. These two examples, flow of current in a sample and vortex imaging, demonstrate the advantages of being able to measure the in-plane component in the former case and the out-of-plane component in the latter.

\begin{figure}
				\begin{center}  
					\includegraphics[width=1\textwidth]{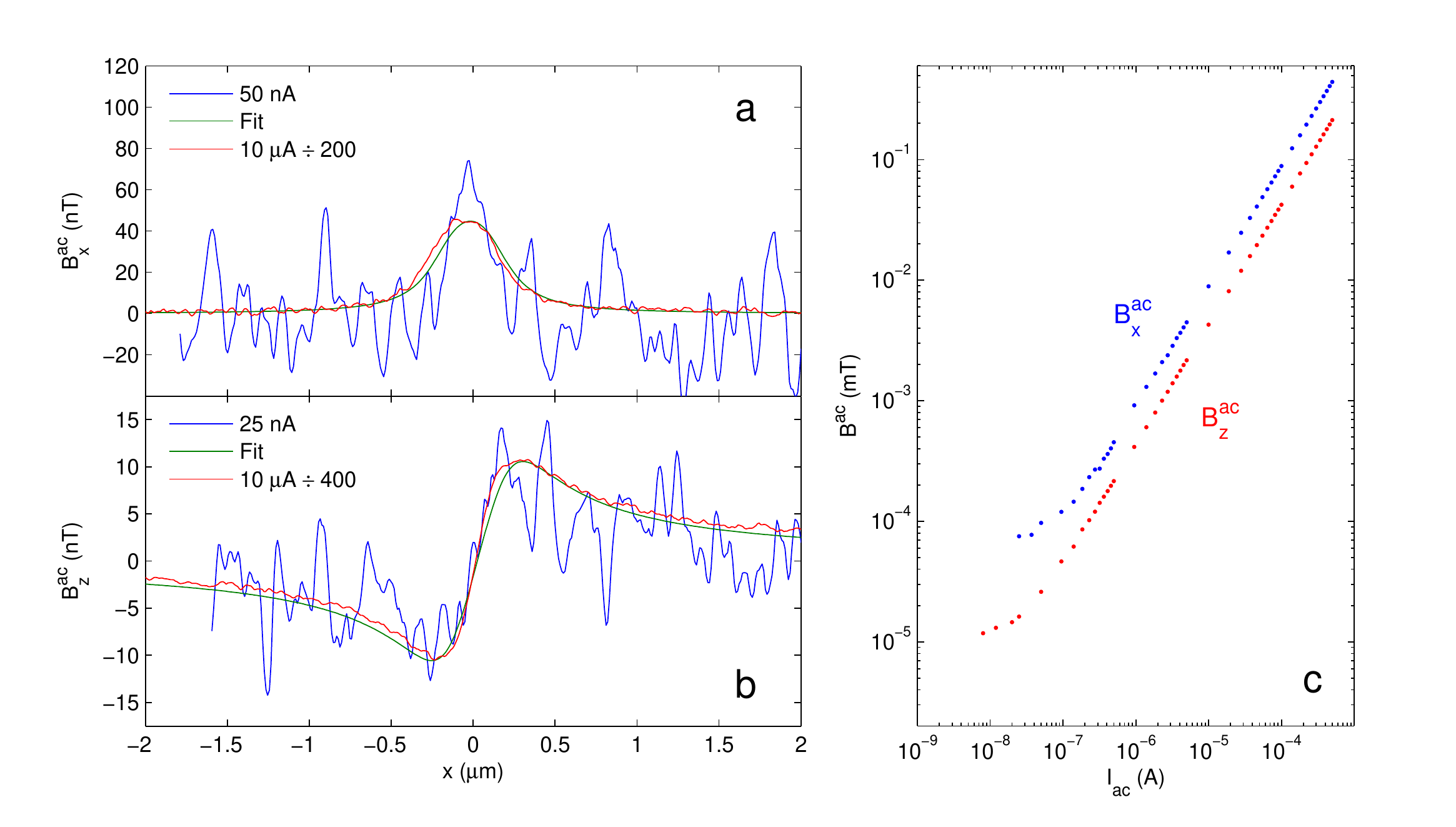}
					\caption[]{\label{profiles}  {\bf Field profile across the center of the nanowire due to an AC current.} (a)-(b) Field profiles $B^{ac}$ across the center of the nanowire. (a) Profiles acquired at a $B_x$-sensitive WP ($\mu_{0} H_{z} = 0\ \mathrm{mT}$, $\mu_{0} H_{x} = 37\ \mathrm{mT}$) while $I_{ac}$ of $50\ \mathrm{nA}$ (blue) and $10\ \upmu \mathrm{A}$ (red, divided by $200$) are applied to the sample. (b) Profiles acquired using a $B_z$-sensitive WP ($\mu_{0} H_{z} = 25\ \mathrm{mT}$, $\mu_{0} H_{x} = -10\ \mathrm{mT}$) while $50\ \mathrm{nA}$ (blue) and $10\ \upmu \mathrm{A}$ (red, divided by $400$) are applied to the sample. Numerical fit to the data is obtained by calculating the magnetic field generated by a uniform distribution of current in an infinitely thin, $350$ nm-wide bridge. The fit assumes two free parameters: the response function of the 3JSOT and the 3JSOT-to-sample distance. The fitted distances were found to be $120$ nm and $150$ nm in the case of the $B_x$- and $B_z$-sensitive WPs, respectively. (c) Plot of the measured amplitude of $B^{ac}_{x}$ (blue) and $B^{ac}_{z}$ (red) vs. the applied $I_{ac}$. (a)-(c) The lock-in amplifier time constant was set to 1 s for currents smaller than 500 nA and no greater than 0.1 s for higher currents.}
					\end{center}
			\end{figure}

Finally, to further demonstrate the ability of the 3JSOT to decouple the two field components and to quantify the sensitivity for imaging of current flow, we measured the $B_z^{ac}$ and $B_x^{ac}$ field profiles across the nanowire, as shown in Fig.\ \ref{profiles} a and b for a relatively large current  of $10\ \upmu \mathrm{A}$ (red line) and for small, yet detectable, currents of $50$ nA and $25$ nA (blue lines) with 1 s integration time per pixel. The theoretical fits (green) show good agreement, demonstrating that we are able to effectively decouple sensitivity to the two components of the field. These measurements were repeated over more than four orders of magnitude in current, from $0.5$ mA down to $10$ nA. The results are summarized in Fig.\ \ref{profiles}c, where the maximum value of the measured field is plotted as a function of $I_{ac}$. For the lowest currents, the measured signal approaches the noise level of our 3JSOT. The lowest detectable currents are about 25 nA and 50 nA for measurements of $B^{ac}_z$ and $B^{ac}_x$ field components, respectively.

In conclusion, we present the fabrication, characterization and imaging capabilities of a unique SQUID that demonstrates a tunable response to both in-plane and out-of-plane fields, while meeting the size and sensitivity standards of state-of-the-art nanoSQUIDs. With a spin sensitivity better than $5\ \frac{\mu_B}{\mathrm{Hz}^{1/2}}$, this versatile tool opens a wide range of possibilities for the study and imaging of nanoscale magnetic systems that were formerly inaccessible.

\section{Acknowledgment}
This work was supported by the European Research Council (ERC advanced grant) and by the Minerva Foundation with funding from the Federal German Ministry for Education and Research. Y.A. acknowledges support by the Azrieli Foundation and by the Fonds Qu\'{e}b\'{e}cois de la Recherche sur la Nature et les Technologies. M.E.H. acknowledges support from the Weston Visiting Professorship program and from a Fulbright Fellowship awarded by the United States-Israel Educational Foundation. E.Z. acknowledges support by the US-Israel Binational Science Foundation (BSF).

\bibliography{3JSOT-bib}

\providecommand{\latin}[1]{#1}
\providecommand*\mcitethebibliography{\thebibliography}
\csname @ifundefined\endcsname{endmcitethebibliography}
  {\let\endmcitethebibliography\endthebibliography}{}
\begin{mcitethebibliography}{36}
\providecommand*\natexlab[1]{#1}
\providecommand*\mciteSetBstSublistMode[1]{}
\providecommand*\mciteSetBstMaxWidthForm[2]{}
\providecommand*\mciteBstWouldAddEndPuncttrue
  {\def\EndOfBibitem{\unskip.}}
\providecommand*\mciteBstWouldAddEndPunctfalse
  {\let\EndOfBibitem\relax}
\providecommand*\mciteSetBstMidEndSepPunct[3]{}
\providecommand*\mciteSetBstSublistLabelBeginEnd[3]{}
\providecommand*\EndOfBibitem{}
\mciteSetBstSublistMode{f}
\mciteSetBstMaxWidthForm{subitem}{(\alph{mcitesubitemcount})}
\mciteSetBstSublistLabelBeginEnd
  {\mcitemaxwidthsubitemform\space}
  {\relax}
  {\relax}

\bibitem[Bending(1999)]{Bending1999}
Bending,~S.~J. \emph{Advances in Physics} \textbf{1999}, \emph{48},
  449--535\relax
\mciteBstWouldAddEndPuncttrue
\mciteSetBstMidEndSepPunct{\mcitedefaultmidpunct}
{\mcitedefaultendpunct}{\mcitedefaultseppunct}\relax
\EndOfBibitem
\bibitem[Grinolds \latin{et~al.}(2013)Grinolds, Hong, Maletinsky, Luan, Lukin,
  Walsworth, and Yacoby]{Grinolds2013}
Grinolds,~M.~S.; Hong,~S.; Maletinsky,~P.; Luan,~L.; Lukin,~M.~D.;
  Walsworth,~R.~L.; Yacoby,~A. \emph{Nature Physics} \textbf{2013}, \emph{9},
  215--219\relax
\mciteBstWouldAddEndPuncttrue
\mciteSetBstMidEndSepPunct{\mcitedefaultmidpunct}
{\mcitedefaultendpunct}{\mcitedefaultseppunct}\relax
\EndOfBibitem
\bibitem[Degen and Home(2011)Degen, and Home]{Degen2011}
Degen,~C.~L.; Home,~J.~P. \emph{Nature Nanotechnology} \textbf{2011}, \emph{6},
  399--400\relax
\mciteBstWouldAddEndPuncttrue
\mciteSetBstMidEndSepPunct{\mcitedefaultmidpunct}
{\mcitedefaultendpunct}{\mcitedefaultseppunct}\relax
\EndOfBibitem
\bibitem[Rugar \latin{et~al.}(2004)Rugar, Budakian, Mamin, and Chui]{Rugar2004}
Rugar,~D.; Budakian,~R.; Mamin,~H.~J.; Chui,~B.~W. \emph{Nature} \textbf{2004},
  \emph{430}, 329--332\relax
\mciteBstWouldAddEndPuncttrue
\mciteSetBstMidEndSepPunct{\mcitedefaultmidpunct}
{\mcitedefaultendpunct}{\mcitedefaultseppunct}\relax
\EndOfBibitem
\bibitem[Tang \latin{et~al.}(2011)Tang, Li, Li, Chi, and Chen]{Tang2011}
Tang,~C.-C.; Li,~M.-Y.; Li,~L.~J.; Chi,~C.~C.; Chen,~J.~C. \emph{Applied
  Physics Letters} \textbf{2011}, \emph{99}, 112107\relax
\mciteBstWouldAddEndPuncttrue
\mciteSetBstMidEndSepPunct{\mcitedefaultmidpunct}
{\mcitedefaultendpunct}{\mcitedefaultseppunct}\relax
\EndOfBibitem
\bibitem[Finkler \latin{et~al.}(2010)Finkler, Segev, Myasoedov, Rappaport,
  Ne'eman, Vasyukov, Zeldov, Huber, Martin, and Yacoby]{Finkler2010}
Finkler,~A.; Segev,~Y.; Myasoedov,~Y.; Rappaport,~M.~L.; Ne'eman,~L.;
  Vasyukov,~D.; Zeldov,~E.; Huber,~M.~E.; Martin,~J.; Yacoby,~A. \emph{Nano
  Lett.} \textbf{2010}, \emph{10}, 1046--1049\relax
\mciteBstWouldAddEndPuncttrue
\mciteSetBstMidEndSepPunct{\mcitedefaultmidpunct}
{\mcitedefaultendpunct}{\mcitedefaultseppunct}\relax
\EndOfBibitem
\bibitem[Finkler \latin{et~al.}(2012)Finkler, Vasyukov, Segev, Ne'eman,
  Lachman, Rappaport, Myasoedov, Zeldov, and Huber]{Finkler2012}
Finkler,~A.; Vasyukov,~D.; Segev,~Y.; Ne'eman,~L.; Lachman,~E.~O.;
  Rappaport,~M.~L.; Myasoedov,~Y.; Zeldov,~E.; Huber,~M.~E. \emph{Review of
  Scientific Instruments} \textbf{2012}, \emph{83}, 073702\relax
\mciteBstWouldAddEndPuncttrue
\mciteSetBstMidEndSepPunct{\mcitedefaultmidpunct}
{\mcitedefaultendpunct}{\mcitedefaultseppunct}\relax
\EndOfBibitem
\bibitem[Hasselbach \latin{et~al.}(2002)Hasselbach, Mailly, and
  Kirtley]{Hasselbach2002}
Hasselbach,~K.; Mailly,~D.; Kirtley,~J.~R. \emph{Journal of Applied Physics}
  \textbf{2002}, \emph{91}, 4432--4437\relax
\mciteBstWouldAddEndPuncttrue
\mciteSetBstMidEndSepPunct{\mcitedefaultmidpunct}
{\mcitedefaultendpunct}{\mcitedefaultseppunct}\relax
\EndOfBibitem
\bibitem[Ronzani \latin{et~al.}(2013)Ronzani, Baillergeau, Altimiras, and
  Giazotto]{Ronzani2013}
Ronzani,~A.; Baillergeau,~M.; Altimiras,~C.; Giazotto,~F. \emph{Applied Physics
  Letters} \textbf{2013}, \emph{103}, 052603\relax
\mciteBstWouldAddEndPuncttrue
\mciteSetBstMidEndSepPunct{\mcitedefaultmidpunct}
{\mcitedefaultendpunct}{\mcitedefaultseppunct}\relax
\EndOfBibitem
\bibitem[W\"{o}lbing \latin{et~al.}(2013)W\"{o}lbing, Nagel, Schwarz, Kieler,
  Weimann, Kohlmann, Zorin, Kemmler, Kleiner, and Koelle]{Wolbing2013}
W\"{o}lbing,~R.; Nagel,~J.; Schwarz,~T.; Kieler,~O.; Weimann,~T.; Kohlmann,~J.;
  Zorin,~A.~B.; Kemmler,~M.; Kleiner,~R.; Koelle,~D. \emph{Applied Physics
  Letters} \textbf{2013}, \emph{102}, 192601\relax
\mciteBstWouldAddEndPuncttrue
\mciteSetBstMidEndSepPunct{\mcitedefaultmidpunct}
{\mcitedefaultendpunct}{\mcitedefaultseppunct}\relax
\EndOfBibitem
\bibitem[Lam \latin{et~al.}(2011)Lam, Clem, and Yang]{Lam2011}
Lam,~S. K.~H.; Clem,~J.~R.; Yang,~W. \emph{Nanotechnology} \textbf{2011},
  \emph{22}, 455501\relax
\mciteBstWouldAddEndPuncttrue
\mciteSetBstMidEndSepPunct{\mcitedefaultmidpunct}
{\mcitedefaultendpunct}{\mcitedefaultseppunct}\relax
\EndOfBibitem
\bibitem[Foley and Hilgenkamp(2009)Foley, and Hilgenkamp]{Foley2009}
Foley,~C.~P.; Hilgenkamp,~H. \emph{Superconductor Science and Technology}
  \textbf{2009}, \emph{22}, 064001\relax
\mciteBstWouldAddEndPuncttrue
\mciteSetBstMidEndSepPunct{\mcitedefaultmidpunct}
{\mcitedefaultendpunct}{\mcitedefaultseppunct}\relax
\EndOfBibitem
\bibitem[Kirtley \latin{et~al.}(1995)Kirtley, Ketchen, Stawiasz, Sun,
  Gallagher, Blanton, and Wind]{Kirtley1995}
Kirtley,~J.~R.; Ketchen,~M.~B.; Stawiasz,~K.~G.; Sun,~J.~Z.; Gallagher,~W.~J.;
  Blanton,~S.~H.; Wind,~S.~J. \emph{Applied Physics Letters} \textbf{1995},
  \emph{66}, 1138--1140\relax
\mciteBstWouldAddEndPuncttrue
\mciteSetBstMidEndSepPunct{\mcitedefaultmidpunct}
{\mcitedefaultendpunct}{\mcitedefaultseppunct}\relax
\EndOfBibitem
\bibitem[Nagel \latin{et~al.}(2011)Nagel, Kieler, Weimann, Wölbing, Kohlmann,
  Zorin, Kleiner, Koelle, and Kemmler]{Nagel2011}
Nagel,~J.; Kieler,~O.~F.; Weimann,~T.; Wölbing,~R.; Kohlmann,~J.;
  Zorin,~A.~B.; Kleiner,~R.; Koelle,~D.; Kemmler,~M. \emph{Applied Physics
  Letters} \textbf{2011}, \emph{99}, 032506\relax
\mciteBstWouldAddEndPuncttrue
\mciteSetBstMidEndSepPunct{\mcitedefaultmidpunct}
{\mcitedefaultendpunct}{\mcitedefaultseppunct}\relax
\EndOfBibitem
\bibitem[Troeman \latin{et~al.}(2007)Troeman, Derking, Borger, Pleikies,
  Veldhuis, and Hilgenkamp]{Troeman2007}
Troeman,~A. G.~P.; Derking,~H.; Borger,~B.; Pleikies,~J.; Veldhuis,~D.;
  Hilgenkamp,~H. \emph{Nano Letters} \textbf{2007}, \emph{7}, 2152--2156\relax
\mciteBstWouldAddEndPuncttrue
\mciteSetBstMidEndSepPunct{\mcitedefaultmidpunct}
{\mcitedefaultendpunct}{\mcitedefaultseppunct}\relax
\EndOfBibitem
\bibitem[Kirtley(2010)]{Kirtley2010}
Kirtley,~J.~R. \emph{Reports on Progress in Physics} \textbf{2010}, \emph{73},
  126501\relax
\mciteBstWouldAddEndPuncttrue
\mciteSetBstMidEndSepPunct{\mcitedefaultmidpunct}
{\mcitedefaultendpunct}{\mcitedefaultseppunct}\relax
\EndOfBibitem
\bibitem[Hao \latin{et~al.}(2008)Hao, Macfarlane, Gallop, Cox, Beyer, Drung,
  and Schurig]{Hao2008}
Hao,~L.; Macfarlane,~J.~C.; Gallop,~J.~C.; Cox,~D.; Beyer,~J.; Drung,~D.;
  Schurig,~T. \emph{Applied Physics Letters} \textbf{2008}, \emph{92},
  192507\relax
\mciteBstWouldAddEndPuncttrue
\mciteSetBstMidEndSepPunct{\mcitedefaultmidpunct}
{\mcitedefaultendpunct}{\mcitedefaultseppunct}\relax
\EndOfBibitem
\bibitem[Levenson-Falk \latin{et~al.}(2013)Levenson-Falk, Vijay, Antler, and
  Siddiqi]{Levenson-Falk2013}
Levenson-Falk,~E.~M.; Vijay,~R.; Antler,~N.; Siddiqi,~I. \emph{Superconductor
  Science and Technology} \textbf{2013}, \emph{26}, 055015\relax
\mciteBstWouldAddEndPuncttrue
\mciteSetBstMidEndSepPunct{\mcitedefaultmidpunct}
{\mcitedefaultendpunct}{\mcitedefaultseppunct}\relax
\EndOfBibitem
\bibitem[Clarke and Braginski(2006)Clarke, and Braginski]{Clarke2006}
Clarke,~J.; Braginski,~A. \emph{The SQUID Handbook: Fundamentals and Technology
  of SQUIDs and SQUID Systems}; Wiley, 2006\relax
\mciteBstWouldAddEndPuncttrue
\mciteSetBstMidEndSepPunct{\mcitedefaultmidpunct}
{\mcitedefaultendpunct}{\mcitedefaultseppunct}\relax
\EndOfBibitem
\bibitem[Koshnick \latin{et~al.}(2008)Koshnick, Huber, Bert, Hicks, Large,
  Edwards, and Moler]{Koshnick2008}
Koshnick,~N.~C.; Huber,~M.~E.; Bert,~J.~A.; Hicks,~C.~W.; Large,~J.;
  Edwards,~H.; Moler,~K.~A. \emph{Applied Physics Letters} \textbf{2008},
  \emph{93}, 243101\relax
\mciteBstWouldAddEndPuncttrue
\mciteSetBstMidEndSepPunct{\mcitedefaultmidpunct}
{\mcitedefaultendpunct}{\mcitedefaultseppunct}\relax
\EndOfBibitem
\bibitem[Hykel \latin{et~al.}(2014)Hykel, Wang, Castellazzi, Crozes, Shaw,
  Schuster, and Hasselbach]{Hykel2014}
Hykel,~D.; Wang,~Z.; Castellazzi,~P.; Crozes,~T.; Shaw,~G.; Schuster,~K.;
  Hasselbach,~K. \emph{Journal of Low Temperature Physics} \textbf{2014},
  \emph{175}, 861--867\relax
\mciteBstWouldAddEndPuncttrue
\mciteSetBstMidEndSepPunct{\mcitedefaultmidpunct}
{\mcitedefaultendpunct}{\mcitedefaultseppunct}\relax
\EndOfBibitem
\bibitem[Vasyukov \latin{et~al.}(2013)Vasyukov, Anahory, Embon, Halbertal,
  Cuppens, Ne'eman, Finkler, Segev, Myasoedov, Rappaport, Huber, and
  Zeldov]{Vasyukov2013}
Vasyukov,~D.; Anahory,~Y.; Embon,~L.; Halbertal,~D.; Cuppens,~J.; Ne'eman,~L.;
  Finkler,~A.; Segev,~Y.; Myasoedov,~Y.; Rappaport,~M.~L.; Huber,~M.~E.;
  Zeldov,~E. \emph{Nature Nanotechnology} \textbf{2013}, \emph{8},
  639--644\relax
\mciteBstWouldAddEndPuncttrue
\mciteSetBstMidEndSepPunct{\mcitedefaultmidpunct}
{\mcitedefaultendpunct}{\mcitedefaultseppunct}\relax
\EndOfBibitem
\bibitem[Sochnikov \latin{et~al.}(2013)Sochnikov, Bestwick, Williams, Lippman,
  Fisher, Goldhaber-Gordon, Kirtley, and Moler]{Sochnikov2013}
Sochnikov,~I.; Bestwick,~A.~J.; Williams,~J.~R.; Lippman,~T.~M.; Fisher,~I.~R.;
  Goldhaber-Gordon,~D.; Kirtley,~J.~R.; Moler,~K.~A. \emph{Nano Letters}
  \textbf{2013}, \emph{13}, 3086--3092\relax
\mciteBstWouldAddEndPuncttrue
\mciteSetBstMidEndSepPunct{\mcitedefaultmidpunct}
{\mcitedefaultendpunct}{\mcitedefaultseppunct}\relax
\EndOfBibitem
\bibitem[Bid \latin{et~al.}(2010)Bid, Ofek, Inoue, Heiblum, Kane, Umansky, and
  Mahalu]{Bid2010}
Bid,~A.; Ofek,~N.; Inoue,~H.; Heiblum,~M.; Kane,~C.~L.; Umansky,~V.; Mahalu,~D.
  \emph{Nature} \textbf{2010}, \emph{466}, 585--590\relax
\mciteBstWouldAddEndPuncttrue
\mciteSetBstMidEndSepPunct{\mcitedefaultmidpunct}
{\mcitedefaultendpunct}{\mcitedefaultseppunct}\relax
\EndOfBibitem
\bibitem[Nowack \latin{et~al.}(2013)Nowack, Spanton, Baenninger, König,
  Kirtley, Kalisky, Ames, Leubner, Brüne, Buhmann, Molenkamp,
  Goldhaber-Gordon, and Moler]{Nowack2013}
Nowack,~K.~C.; Spanton,~E.~M.; Baenninger,~M.; König,~M.; Kirtley,~J.~R.;
  Kalisky,~B.; Ames,~C.; Leubner,~P.; Brüne,~C.; Buhmann,~H.;
  Molenkamp,~L.~W.; Goldhaber-Gordon,~D.; Moler,~K.~A. \emph{Nature Materials}
  \textbf{2013}, \emph{12}, 787--791\relax
\mciteBstWouldAddEndPuncttrue
\mciteSetBstMidEndSepPunct{\mcitedefaultmidpunct}
{\mcitedefaultendpunct}{\mcitedefaultseppunct}\relax
\EndOfBibitem
\bibitem[Tiemann \latin{et~al.}(2012)Tiemann, Gamez, Kumada, and
  Muraki]{Tiemann2012}
Tiemann,~L.; Gamez,~G.; Kumada,~N.; Muraki,~K. \emph{Science} \textbf{2012},
  \emph{335}, 828--831\relax
\mciteBstWouldAddEndPuncttrue
\mciteSetBstMidEndSepPunct{\mcitedefaultmidpunct}
{\mcitedefaultendpunct}{\mcitedefaultseppunct}\relax
\EndOfBibitem
\bibitem[Kuemmeth \latin{et~al.}(2008)Kuemmeth, Ilani, Ralph, and
  McEuen]{Kuemmeth2008}
Kuemmeth,~F.; Ilani,~S.; Ralph,~D.~C.; McEuen,~P.~L. \emph{Nature}
  \textbf{2008}, \emph{452}, 448--452\relax
\mciteBstWouldAddEndPuncttrue
\mciteSetBstMidEndSepPunct{\mcitedefaultmidpunct}
{\mcitedefaultendpunct}{\mcitedefaultseppunct}\relax
\EndOfBibitem
\bibitem[Kalisky \latin{et~al.}(2012)Kalisky, Bert, Bell, Xie, Sato, Hosoda,
  Hikita, Hwang, and Moler]{Kalisky2012}
Kalisky,~B.; Bert,~J.~A.; Bell,~C.; Xie,~Y.; Sato,~H.~K.; Hosoda,~M.;
  Hikita,~Y.; Hwang,~H.~Y.; Moler,~K.~A. \emph{Nano Letters} \textbf{2012},
  \emph{12}, 4055--4059\relax
\mciteBstWouldAddEndPuncttrue
\mciteSetBstMidEndSepPunct{\mcitedefaultmidpunct}
{\mcitedefaultendpunct}{\mcitedefaultseppunct}\relax
\EndOfBibitem
\bibitem[Romans \latin{et~al.}(2010)Romans, Osley, Young, Warburton, and
  Li]{Romans2010}
Romans,~E.~J.; Osley,~E.~J.; Young,~L.; Warburton,~P.~A.; Li,~W. \emph{Applied
  Physics Letters} \textbf{2010}, \emph{97}, 222506\relax
\mciteBstWouldAddEndPuncttrue
\mciteSetBstMidEndSepPunct{\mcitedefaultmidpunct}
{\mcitedefaultendpunct}{\mcitedefaultseppunct}\relax
\EndOfBibitem
\bibitem[Chiarello \latin{et~al.}(2008)Chiarello, Castellano, Torrioli,
  Poletto, Cosmelli, Carelli, Balashov, Khabipov, and Zorin]{Chiarello2008}
Chiarello,~F.; Castellano,~M.~G.; Torrioli,~G.; Poletto,~S.; Cosmelli,~C.;
  Carelli,~P.; Balashov,~D.~V.; Khabipov,~M.~I.; Zorin,~A.~B. \emph{Applied
  Physics Letters} \textbf{2008}, \emph{93}, 042504\relax
\mciteBstWouldAddEndPuncttrue
\mciteSetBstMidEndSepPunct{\mcitedefaultmidpunct}
{\mcitedefaultendpunct}{\mcitedefaultseppunct}\relax
\EndOfBibitem
\bibitem[Mart\'{i}nez-P\'{e}rez and Giazotto(2013)Mart\'{i}nez-P\'{e}rez, and
  Giazotto]{Martinez-Perez2013}
Mart\'{i}nez-P\'{e}rez,~M.~J.; Giazotto,~F. \emph{Applied Physics Letters}
  \textbf{2013}, \emph{102}, 092602\relax
\mciteBstWouldAddEndPuncttrue
\mciteSetBstMidEndSepPunct{\mcitedefaultmidpunct}
{\mcitedefaultendpunct}{\mcitedefaultseppunct}\relax
\EndOfBibitem
\bibitem[Ronzani \latin{et~al.}(2014)Ronzani, Altimiras, and
  Giazotto]{Ronzani2014}
Ronzani,~A.; Altimiras,~C.; Giazotto,~F. \emph{Applied Physics Letters}
  \textbf{2014}, \emph{104}, 032601\relax
\mciteBstWouldAddEndPuncttrue
\mciteSetBstMidEndSepPunct{\mcitedefaultmidpunct}
{\mcitedefaultendpunct}{\mcitedefaultseppunct}\relax
\EndOfBibitem
\bibitem[Zeldov \latin{et~al.}(1994)Zeldov, Clem, McElfresh, and
  Darwin]{Zeldov1994}
Zeldov,~E.; Clem,~J.~R.; McElfresh,~M.; Darwin,~M. \emph{Phys. Rev. B}
  \textbf{1994}, \emph{49}, 9802--9822\relax
\mciteBstWouldAddEndPuncttrue
\mciteSetBstMidEndSepPunct{\mcitedefaultmidpunct}
{\mcitedefaultendpunct}{\mcitedefaultseppunct}\relax
\EndOfBibitem
\bibitem[Schuster \latin{et~al.}(1995)Schuster, Kuhn, Brandt, Indenbom,
  Kläser, Müller-Vogt, Habermeier, Kronmüller, and Forkl]{Schuster1995}
Schuster,~T.; Kuhn,~H.; Brandt,~E.~H.; Indenbom,~M.~V.; Kläser,~M.;
  Müller-Vogt,~G.; Habermeier,~H.-U.; Kronmüller,~H.; Forkl,~A. \emph{Phys.
  Rev. B} \textbf{1995}, \emph{52}, 10375--10389\relax
\mciteBstWouldAddEndPuncttrue
\mciteSetBstMidEndSepPunct{\mcitedefaultmidpunct}
{\mcitedefaultendpunct}{\mcitedefaultseppunct}\relax
\EndOfBibitem
\bibitem[Dinner \latin{et~al.}(2005)Dinner, Beasley, and Moler]{Dinner2005}
Dinner,~R.~B.; Beasley,~M.~R.; Moler,~K.~A. \emph{Review of Scientific
  Instruments} \textbf{2005}, \emph{76}, 103702\relax
\mciteBstWouldAddEndPuncttrue
\mciteSetBstMidEndSepPunct{\mcitedefaultmidpunct}
{\mcitedefaultendpunct}{\mcitedefaultseppunct}\relax
\EndOfBibitem
\end{mcitethebibliography}

\end{document}